\documentclass[aps,prl,reprint,groupedaddress]{revtex4-1} 
\usepackage[colorlinks=true,urlcolor=blue]{hyperref}
\usepackage{amssymb,amsmath,amsfonts}
\usepackage{color}
\usepackage{graphicx}
\usepackage{array}
\usepackage{epsfig}
\usepackage{epstopdf}
\usepackage{caption}
\usepackage{subcaption}
\usepackage{epsfig}
\usepackage{epstopdf}
\usepackage{eso-pic}
\usepackage{float}

\def\clock{{\count0=\time
           \divide\count0 60
           \ifnum\count0<10 0\fi\the\count0
           \multiply\count0 -60 \advance\count0 \time
           :\ifnum\count0<10 0\fi \the\count0
         }}
\newcommand{\timestamp}{{\small\vbox{\hbox{\tt\jobname.tex}
\hbox{\the\day/\the\month/\the\year, \clock}}}}

\newcommand{\be}{\begin{eqnarray}}
\newcommand{\ee}{\end{eqnarray}}
\newcommand{\beq}{\begin{eqnarray}}
\newcommand{\eeq}{\end{eqnarray}}

\newcommand{\beqa}{\begin{eqnarray}}
\newcommand{\eeqa}{\end{eqnarray}}

\let\oldsqrt\sqrt
\def\sqrt{\mathpalette\DHLhksqrt}
\def\DHLhksqrt#1#2{%
\setbox0=\hbox{$#1\oldsqrt{#2\,}$}\dimen0=\ht0
\advance\dimen0-0.2\ht0
\setbox2=\hbox{\vrule height\ht0 depth -\dimen0}%
{\box0\lower0.4pt\box2}}

\newcommand\BackgroundPic{%
\put(0,0){%
\parbox[b][\paperheight]{\paperwidth}{%
\vfill
\centering
\includegraphics[width=\paperwidth,height=\paperheight,%
keepaspectratio]{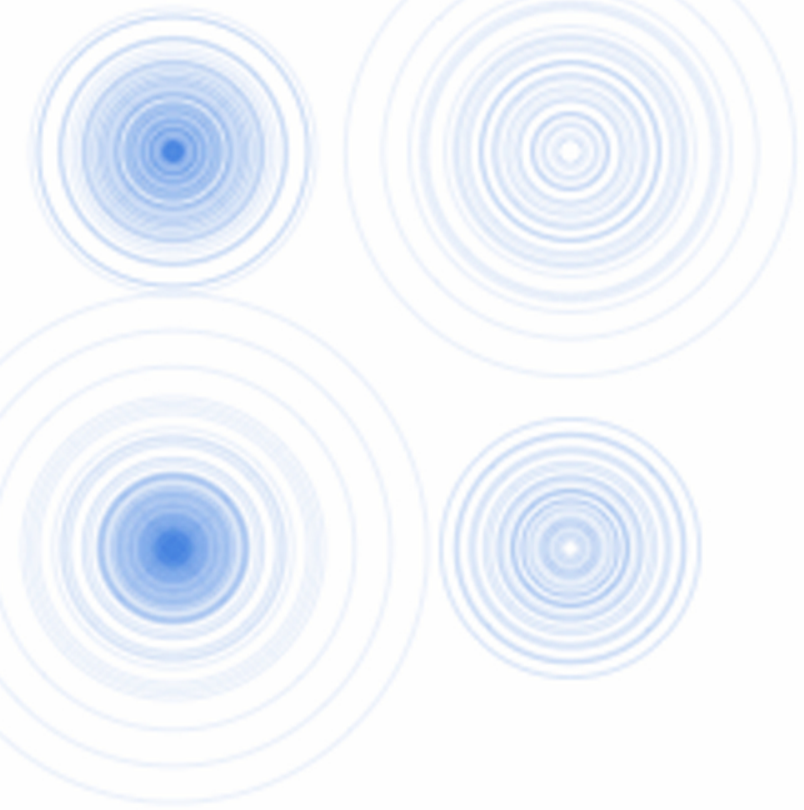}%
\vfill
}}}

\begin{document}
\AddToShipoutPicture*{\BackgroundPic}

\title{Black Holes and Biophysical (Mem)-branes}

\author{Jay Armas$^{a}$ and Troels Harmark$^{b}$}
\email[]{jay@itp.unibe.ch, harmark@nbi.dk}
\affiliation{$^{a}$Albert Einstein Center for Fundamental Physics, University of Bern \\ Sidlerstrasse 5, 3012 Bern}
\affiliation{$^{b}$The Niels Bohr Institute, University of Copenhagen \\
Blegdamsvej 17, 2100 Copenhagen ¯}



\begin{abstract}
We argue that the effective theory describing the long-wavelength dynamics of black branes is the same effective theory that describes the dynamics of biophysical membranes. We improve the phase structure of higher-dimensional black rings by considering finite thickness corrections in this effective theory, showing a striking agreement between our analytical results and recent numerical constructions while simultaneously drawing a parallel between gravity and the effective theory of biophysical membranes.

\end{abstract}


\maketitle

\noindent\textbf{Introduction.}
In the past three years many remarkable properties of higher-dimensional black holes were uncovered, leading to a new perspective on black holes as materials describable by effective theories of continuous media. In particular, in a long-wavelength regime, besides having fluid-like properties, they can behave as elastic (mem)-branes if bent and exhibit piezoelectric behaviour if charged and flexed. In this regime they are characterised by a set of transport coefficients such as shear and bulk viscosities \cite{Policastro:2001yc, Bhattacharyya:2008jc}, the Young modulus \cite{Armas:2011uf} and the piezoelectric moduli \cite{Armas:2012ac}, which can be directly measured from gravity.

Fluid and (mem)-brane elastic behaviour of a physical system can be described by a single unified framework of hydrodynamics on embedded surfaces \cite{Armas:2013hsa, Armas:2013goa}. When applied to black holes and black branes, this framework is commonly known as the blackfold approach \cite{Emparan:2009cs, Emparan:2009at}, which has enabled a systematic scan of new horizon topologies in higher-dimensions \cite{Emparan:2009vd} and the construction of new approximate solutions, such as black rings in arbitrary space-time dimensions \cite{Emparan:2007wm}.

The blackfold approach, as originally developed in \cite{Emparan:2009cs, Emparan:2009at}, consists in applying generic long wavelength perturbations of neutral asymptotically flat black $p$-brane metrics and describing its dynamics via an effective theory. In this case, to leading order in the perturbation, the black branes are described by a set of world volume fields composed of the induced metric $\gamma_{ab}$, the local boost velocity $u^{a}$ and the brane thickness $r_0$. Focusing on the stationary sector, the dynamics of these branes can be described by the free energy functional of a fluid living on an elastic brane. To next order in the perturbation, the black brane metric is corrected by terms proportional to derivatives of the fields $\gamma_{ab},u^{a},r_0$ which include, as we will explain shortly, terms proportional to the extrinsic curvature tensor ${K_{ab}}^{i}$, the Riemann curvature tensor of the world volume, or the vorticity of the fluid. \footnote{Derivatives of the thickness $r_0$ can be exchanged by derivatives of the fluid velocity \cite{Armas:2013hsa, Armas:2013goa}. Furthermore, the black branes can also be rotating on transverse directions in which case one also must take into account contributions due the extrinsic twist potential ${\omega_{a}}^{ij}$ \cite{Armas:2013hsa, Armas:2013goa}.}

The dynamics of elastic membranes has also taken an important role in the context of biophysics, beginning with the early work of Helfrich \cite{Helfrich1973} and Canham \cite{Canham197061} in the 60's. They showed that the bending deformations of elastic membranes played a crucial role in the understanding of the shape of red blood cells. In detail these arise by adding the Helfrich-Canham bending energy, consisting of adding a piece of the form
\beq \label{HC}
\mathcal{F}_{\text{HC}}[X^{\mu}(\sigma^{a})]=\alpha \int_{A}dA \thinspace K^{2}~~,
\eeq
to the free energy of a biophysical membrane. Here $X^{\mu}(\sigma^a)$ describes the position of the membrane in its involving space, $\sigma^{a}$ are a set of coordinates on the membrane, $A$ its area and $K$ its mean extrinsic curvature. The addition of such terms \eqref{HC} has lead to an exhaustive study of this fluid-elastic system \cite{doi:10.1080/00018739700101488}.

Black holes in the long-wavelength regime are described by an effective fluid living on a (mem)-brane \cite{Emparan:2009at}. Focusing on stationary configurations of this effective fluid, and hence on stationary black hole solutions, it is possible, using equilibrium partition function techniques for hydrodynamics \cite{Banerjee:2012iz, Jensen:2012jh}, to construct an effective theory for fluids living on (mem)-branes in a derivative expansion. The resulting effective free energy to second order in the expansion, assuming no transverse spin and codimension higher than one, can be written as \cite{Armas:2013hsa}
\beq \label{free}
\begin{split}
\mathcal{F}[X^{\mu}(\sigma^{a})]&=-\int_{\mathcal{B}_{p}}\!\!dV_{(p)}R_0\Big(P +\upsilon_1 \omega^{ab}\omega_{ab}+ \upsilon_2 \mathcal{R} \\
&+\upsilon_3 u^{a}u^{b}\mathcal{R}_{ab} +\lambda_1 K^{i}K_{i}+\lambda_2 K^{abi}K_{abi} \\
&+\lambda_3 u^{a}u^{b}{K_{a}}^{ci}K_{bci}\Big)~~, \\
\end{split}
\eeq
generalizing the Helfrich-Canham bending energy \eqref{HC} to Lorentzian membranes of arbitrary codimension and accounting for the fact that the fluid living on the membrane can be in a stationary motion. In Eq.~\eqref{free} we have introduced the fluid pressure $P$, the fluid velocity $u^{a}$ and a set of transport coefficients $\upsilon_1,\upsilon_2,\upsilon_3,\lambda_1,\lambda_2,\lambda_3$ which only depend on the local fluid temperature and chemical potentials. Furthermore, $\mathcal{B}_{p}$ denotes the $p$-dimensional submanifold of the $(p+1)$-dimensional world volume $\mathcal{W}_{p+1}$ of the brane, $dV_{(p)}$ its infinitesimal volume and $R_0$ the redshift factor defined via $d^{p+1}\sigma \sqrt{-\gamma}=R_0 d\sigma^{0} dV_{(p)}$ with $\gamma$ being the determinant of the induced metric $\gamma_{ab}={u_{a}}^{\mu}{u_{b}}^{\nu}g_{\mu\nu}$, where ${u_{a}}^{\mu}=\partial_{a}X^{\mu}$ and $g_{\mu\nu}$ the $D$-dimensional ambient metric.

Besides the bending energy \eqref{HC}, the effective theory \eqref{free} accounts for corrections describing the response of the fluid due to a non-zero vorticity $\omega_{ab}$, the possibility of non-zero world volume Ricci scalar $\mathcal{R}$ and tensor $\mathcal{R}_{ab}$ as well as possible elastic (mem)-brane effects due to a non-vanishing extrinsic curvature tensor ${K_{ab}}^{i}={n^{i}}_{\mu}\nabla_{a}{u_{b}}^{\mu}$, where ${n^{i}}_{\mu}$ is an orthogonal projector to the brane, and a non-vanishing mean extrinsic curvature $K^{i}=\gamma^{ab}{K_{ab}}^{i}$. However, despite the fact that the effective theory \eqref{free} is characterised by six transport coefficients, for the case of black rings, as we shall see, only one is necessary and the only correction in \eqref{free} is the generalisation to arbitrary codimension of the bending energy \eqref{HC}.

We note that, in the first blackfold works \cite{Emparan:2009cs, Emparan:2009at}, the effective theory \eqref{free} has only been constructed to leading order, meaning that only the first term proportional to $P$ has been considered, where $P$ is the pressure measured from the black brane metric. To higher orders, as mentioned previously, one needs to measure the set of transport coefficients $\upsilon_1,\upsilon_2,\upsilon_3,\lambda_1,\lambda_2,\lambda_3$. At the present moment, only the transport coefficients $\lambda_1,\lambda_2,\lambda_3$ have been measured from the first order corrected metric \cite{Armas:2011uf, Camps:2012hw} while $\upsilon_1,\upsilon_2,\upsilon_3$ have not. However, these are not required for the purpose of the application that we will describe in this note.

The type of corrections in \eqref{free} are present in large classes of higher-dimensional black holes and our aim in this brief note is to exemplify how the effective theory \eqref{free} can be used to obtain generic features about large classes of black holes. In particular we will see that, similarly to the fact that \eqref{HC} allowed for biophysical membrane configurations with increased surface area, the elastic effects of \eqref{free} result in an increased area (entropy) of black rings for a given value of their angular momenta. These results are in striking agreement with those obtained recently by solving Einstein's equations numerically, therefore confirming the veracity of the effective theory \eqref{free} for higher-dimensional black holes and hence drawing a novel parallel between effective theories of black holes and the effective theories of biophysical membranes.

Before we continue on to describing this effective theory in more detail we would like to mention some of the relations between the effective theory \eqref{free} and those well studied in the context of biophysical membranes. The theory that we have presented in \eqref{free} was assumed to be relativistic and for branes of codimension higher than one. Biophysical membranes are non-relativistic two-dimensional membranes embedded in three-dimensional Euclidean space. In order to obtain the non-relativistic version of \eqref{free} one should simply consider an Euclidean background metric $g_{\mu\nu}$ and take the world volume indices $a,b,c...$ to be only spatial indices. Furthermore, because they are codimension-1 branes, it means that the extrinsic curvature tensor only has one direction and hence a term proportional to $K$, where we have omitted the index $i$, can be added to \eqref{free} as well as terms proportional to worldvolume derivatives of $K$ \cite{0305-4470-36-23-301, Armas:2013hsa}. A recent review of different models for biophysical membranes and their usage can be found in \cite{Tu201466}. Despite the fact that the black branes that we will consider are generically of codimension higher than one, the black ring configuration that we will analyse in detail is described by a two-dimensional membrane with only one extrinsic curvature component and hence effectively behaving like a fluid membrane of codimension-1. Furthermore, we should also note that for generic black branes one should also include back reaction effects, which are not present in the case of biophysical membranes. This however, as we will mention below, can be ignored if the black brane codimension is higher than three. We will now proceed and describe the dynamics and thermodynamics that are obtained from \eqref{free}.


\noindent\textbf{(Mem)-brane dynamics and thermodynamics.}
The dynamics of the (mem)-branes described by the effective theory \eqref{free} can be obtained by varying \eqref{free} with respect to the induced metric and extrinsic curvature tensor. The resulting equations of motion are \cite{Armas:2013hsa}
\beq \label{eq1}
\nabla_{a}T^{ab}={u^{b}}_{\mu}\nabla_{a}\nabla_{b}\mathcal{D}^{ab\mu}+\mathcal{D}^{aci}{R^{b}}_{aic} ~~,
\eeq
\beq \label{eq2}
T^{ab}{K_{ab}}^{i}={n^{i}}_{\mu}\nabla_a\nabla_b\mathcal{D}^{ab\mu}+\mathcal{D}^{abj}{R^{i}}_{ajb}~~,
\eeq
where $R_{\mu\nu\lambda\rho}$ is the background Riemann tensor and where we have defined
\beq \label{td}
T^{ab}=-\frac{2}{\sqrt{-\gamma}}\frac{\delta \mathcal{\mathcal{F}}}{\delta\gamma_{ab}}~~,~~{\mathcal{D}^{ab}}_{i}=-\frac{1}{\sqrt{-\gamma}}\frac{\delta \mathcal{\mathcal{F}}}{{\delta K_{ab}}^{i}}~~.
\eeq
Throughout this note, we use the indices $a,b,..$ to label the $p+1$ directions along the brane and $i,j,k...$ to label the $n+2$ directions transverse to the brane. The greek indices $\mu,\nu...$ label generic space-time indices along the $D=n+p+3$ coordinates. In Eq.~\eqref{td} we have introduced the stress-tensor of the brane $T^{ab}$ and its bending moment $\mathcal{D}^{abi}$ with the form 
\beq
\mathcal{D}^{abi}=\mathcal{Y}^{abcd}{K_{cd}}^{i}~~,
\eeq
where the tensor structure $\mathcal{Y}^{abcd}$ is the Young modulus of the (mem)-brane and reads
\beq
\mathcal{Y}^{abcd}=2\left(\lambda_1\gamma^{ab}\gamma^{cd}+\lambda_2\gamma^{a(c}\gamma^{d)b}+\lambda_3u^{(a}\gamma^{b)(c}u^{d)}\right)~,~~
\eeq 
encoding the Hookean response of the (mem)-brane to bending deformations. 

Eq.~\eqref{eq1} is a non-trivial identity which encodes stress-energy conservation while Eq.~\eqref{eq2} encodes the non-trivial elastodynamics of the fluid. Eqs.~\eqref{eq1}-\eqref{eq2} are relativistic generalisations of the equations of motion for thin elastic membranes of arbitrary codimension \cite{Armas:2013hsa}. In the case where bending deformations are ignored, i.e. $\mathcal{D}^{abi}=0$, they have been shown to arise as constraint equations directly from Einstein equations by bending black branes \cite{Emparan:2007wm, Camps:2012hw}.

The theory described by \eqref{free} is an effective theory to second order in a derivative expansion. As all such effective theories, it enjoys of a perturbative symmetry \cite{Armas:2013hsa, Armas:2013goa}, or in other words, it is invariant under certain field redefinitions. The free energy \eqref{free} and the equations of motion \eqref{eq1}-\eqref{eq2} are indeed invariant under these field redefinitions, which displace the world volume of the brane $\mathcal{W}_{p+1}$ by a small amount $\tilde\varepsilon^i$, i.e.,
\beq \label{fr}
X^{i}(\sigma^{a})\to X^{i}(\sigma^a)+\tilde\varepsilon^{i}~~,
\eeq
inducing the transformations \cite{Armas:2013goa}
\beq
\delta T^{ab}=T^{ab}_{(0)}\tilde\varepsilon^{i}K_{i}-E^{abcd}{K_{cd}}^{i}\tilde\varepsilon_{i}~~,~~\delta \mathcal{D}^{abi}=T_{(0)}^{ab}\tilde\varepsilon^{i}~~,~~
\eeq
where we have split the stress-energy tensor \eqref{td} into a perfect fluid (leading order) part $T^{ab}_{(0)}$ and a second order correction $\Pi^{ab}_{(2)}$ proportional to the transport coefficients introduced in \eqref{free} according to $T^{ab}=T^{ab}_{(0)}+\Pi^{ab}_{(2)}$. $E^{abcd}$ is the leading order elasticity tensor, defined as \cite{Armas:2012jg}
\beq \label{0st}
T^{ab}_{(0)}=P\gamma^{ab}-P'\textbf{k}u^{a}u^{b}~~,~~E^{abcd}=-2\frac{\partial T^{ab}_{(0)}}{\partial \gamma_{cd}}~~,
\eeq
where we have used the fact that for stationary fluid configurations the fluid velocity must be aligned with a world volume Killing vector field $\textbf{k}^{a}$ such that $u^{a}=\textbf{k}^{a}/\textbf{k}$ with $\textbf{k}=|-\gamma_{ab}\textbf{k}^{a}\textbf{k}^{b}|$ and that the local temperature depends on the global temperature $T$ via $\mathcal{T}=T/\textbf{k}$. The prime in Eq.~\eqref{0st} denotes the derivative with respect to $\textbf{k}$.

Generically, for any stationary fluid configuration we can write the Killing vector field $\textbf{k}^{a}$ as
\beq
\textbf{k}^{a}=\xi^{a}+\Omega^{(b)}\chi_{(b)}^{a}~~,
\eeq
where $\xi^{a}$ denotes a timelike Killing vector field associated with space-time translations, $\chi_{(b)}^{a}$ denotes a set of space-like Killing vectors fields associated with rotational symmetries and $\Omega^{(b)}$ the corresponding angular velocities. In terms of these Killing vectors we can write all conserved quantities associated with the fluid configuration to all orders in a derivative expansion. In order to do so, we note that , focusing on uncharged configurations, we can write the free energy \eqref{free} as
\beq
\mathcal{F}=M-TS-\Omega^{(a)}J_{(a)}~~,
\eeq
where $M$ represents the energy of the fluid, $S$ the entropy and $J^{(a)}$ the angular momentum associated with each rotational isometry. Requiring the variation of the free energy \eqref{free} to vanish leads to the first law of thermodynamics
\beq \label{1stlaw}
dM=TdS+\Omega^{(a)}dJ_{(a)}~~,
\eeq
and hence, defining $\mathcal{F}=-\int_{\mathcal{B}_p}dV_{(p)}R_0\mathcal{L}$, we obtain the thermodynamic expressions 
\beq \label{t1}
M=-\int_{\mathcal{B}_{p}}dV_{(p)}R_0\left(\mathcal{L}+\xi^{a}\frac{\partial\mathcal{L}}{\partial \textbf{k}^{a}}\right)~~,
\eeq
\beq
J_{(a)}=-\left(\frac{\partial \mathcal{F}}{\partial \Omega_{(a)}}\right)_{T}=\int_{\mathcal{B}_{p}}dV_{(p)}R_0 \chi^{b}_{(a)}\frac{\partial\mathcal{L}}{\partial \textbf{k}^{b}}~~,
\eeq
\beq \label{t3}
S=-\left(\frac{\partial \mathcal{F}}{\partial T}\right)_{\Omega_{(a)}}=-\frac{1}{T}\int_{\mathcal{B}_{p}}dV_{(p)}R_0 \textbf{k}^{a}\frac{\partial\mathcal{L}}{\partial \textbf{k}^{a}}~~.
\eeq
The proof of these formulas and the equivalence of \eqref{1stlaw} with the equations of motion \eqref{eq1}-\eqref{eq2} is presented in \cite{Armas:2014rva}.

\noindent\textbf{Biophysical membrane effect in black rings.}
We now wish to apply the effective theory \eqref{free} to the case of higher-dimensional black rings in asymptotically flat space in the thin ring limit. This corresponds to finite thickness corrections in the blackfold approach going beyond the infinitely thin ring approximation of \cite{Emparan:2007wm}. In such case the leading order pressure $P$ reads \cite{Emparan:2009at}
\beq
P=-\frac{\Omega_{(n+1)}}{16\pi G}r_0^{n}~~,~~r_0=\frac{n}{4\pi T}\textbf{k}~~,
\eeq
where $r_0$ is the horizon thickness of the black brane and $\Omega_{(n+1)}$ the volume of an $(n+1)$-sphere. We place the ring in flat space-time parametrised as
\beq
ds^2=-dt^2+dr^2+r^2d\psi^2+\sum_{i=1}^{D-3}dx_{i}^2~~,
\eeq 
by choosing the embedding functions $X^{t}=\tau~,~X^{r}=R~,~X^{\psi}=\phi~,~X^{i}=0$. With this choice the world volume is manifestly flat and hence terms proportional to world volume curvatures in \eqref{free} vanish. Furthermore the Killing vector field is chosen such that $\textbf{k}^{a}=\partial_\tau+\Omega\partial_\phi$ for constant $\Omega$, leading to a vanishing vorticity $\omega_{ab}$. Using Gauss-Codazzi equations, the leading order equation of motion and the field redefinition \eqref{fr}, black rings are only described by one single transport coefficient to second order \cite{Armas:2013hsa, Armas:2013goa},
\beq
\tilde \lambda_1=\frac{\Omega_{(n+1)}}{16\pi G}r_0^{n+2}\frac{(n+1)(3n+4)}{2n^2(n+2)}\xi(n)~~,
\eeq
where $\tilde\lambda_1=\lambda_1+\lambda_2+(1/n)\lambda_3$ and
\beq
\xi(n)=\frac{n\tan(\pi/n)}{\pi}\frac{\Gamma\left(\frac{n+1}{n}\right)^{4}}{\Gamma\left(\frac{n+2}{n}\right)^{2}}~,~n\ge3~~.
\eeq
Since that the only non-vanishing extrinsic curvature component is ${K_{\phi\phi}}^{r}=-R$, the free energy \eqref{free} for black rings is thus 
\beq \label{fbr}
\mathcal{F}[R]=-2\pi R\left(P+\tilde\lambda_1 K^{i}K_{i}\right)~~,
\eeq
where $K^{i}K_{i}=R^{-2}$. Varying this free energy with respect to $R$ and solving the equation of motion \eqref{eq2} leads to an equilibrium condition for the angular velocity \footnote{This equilibrium condition is related to the one found previously in \cite{Armas:2013hsa} via the field redefinition $R\to R+ \frac{(n+1)(n+4)}{n^2(n+2)}\xi(n)r_0\varepsilon$.}
\beq
\Omega=\bar{\Omega}\left(1+\frac{(n+1)(3n+4)}{2n^2(n+2)}\xi(n)\varepsilon^2\right)~~,
\eeq
where we have introduced the dimensionless order parameter $\varepsilon=r_0/R\ll 1$ and the leading order result
\beq
\bar{\Omega}=\frac{1}{\sqrt{n+1}R}~~,
\eeq
previously obtained in \cite{Emparan:2007wm}. Using the thermodynamic expressions \eqref{t1}-\eqref{t3}, which are now interpreted as the mass, angular momentum and entropy of the black hole, we obtain
\beq \label{T1}
\frac{M}{\Omega_{(n+1)}(n\!+\!2)}=\frac{r_0^n}{8G}R\left(1-\frac{(n\!+\!1)(3n\!+\!4)\xi\varepsilon^2}{2n^2(n\!+\!2)}\right),~
\eeq
\beq
\frac{J}{\Omega_{(n+1)}}=\frac{r_0^n}{8G}R^2\sqrt{n\!+\!1}\left(1+\mathcal{O}\left(\varepsilon^4\right)\right)~~,
\eeq
\beq\label{T3}
\frac{S n^{\frac{1}{2}}}{\Omega_{(n+1)}(n\!+\!1)^{\frac{1}{2}}}\!=\!\frac{\pi r_0^{n+1}}{2G}R\!\left(\!\!1\!-\!\frac{(n\!+\!1)^2(3n\!+\!4)\xi\varepsilon^2}{2n^3(n\!+\!2)}\!\right)\!,
\eeq
where we have omitted the dependence of $\xi$ on $n$. These thermodynamic expressions \eqref{T1}-\eqref{T3} are not invariant under field redefinitions \eqref{fr}. In order to present invariant results we introduce the reduced angular momentum, area, angular velocity and temperature as in \cite{Emparan:2007wm},
\beq
j^{n+1}=c_j \frac{J^{n+1}}{GM^{n+2}}~~,~~a_{\text{H}}^{n+1}=4^{n+1}c_a\frac{S^{n+1}}{(GM)^{n+2}}~~,
\eeq
\beq
\omega_{\text{H}}=c_\omega \Omega(GM)^{\frac{1}{n+1}}~~,~~t_{\text{H}}=c_t T (GM)^{\frac{1}{n+1}}~~,
\eeq
with
\beq
c_j=\frac{(16\pi)^{n+1}}{2^{n+4}n^{\frac{n+1}{2}}}c_a=\frac{\Omega_{(n+1)}}{2^{n+5}}\frac{(n+2)^{n+2}}{(n+1)^{\frac{n+1}{2}}}~~,
\eeq
\beq
c_{\omega}=\frac{\sqrt{n}}{4\pi}8^{\frac{1}{n+1}}c_t=\sqrt{n+1}\left(\frac{n+2}{16}\Omega_{(n+1)}\right)^{-\frac{1}{n+1}}.~~
\eeq
Using the freedom given by the field redefinition \eqref{fr} we choose a gauge for which $j$ does not receive any corrections to order $\varepsilon^2$, this is done by performing the transformation $R\to R-\frac{(3n+4)}{2n^2}\xi(n)r_0\varepsilon$. With this we obtain the improved phase structure of higher-dimensional black rings in dimensions $D\ge7$,
\beq \label{a1}
a_{\text{H}}(j)=\frac{2^{\frac{n-2}{n(n+1)}}}{j^{\frac{1}{n}}}\left(1+\frac{(n+1)(3n+4)}{2^{\frac{3n+4}{n}}n^3(n+2)}\frac{\xi(n)}{j^{\frac{2(n+1)}{n}}}\right)~,~~~
\eeq
\beq
\omega_{\text{H}}(j)=\frac{1}{2j}\left(1+\frac{(n+1)(3n+4)}{2^{\frac{2(n+2)}{n}}n^2(n+2)}\frac{\xi(n)}{j^{\frac{2(n+1)}{n}}}\right)~~,~~~
\eeq
\beq\label{a3}
t_{\text{H}}(j)=\frac{nj^{\frac{1}{n}}}{2^{\frac{n-2}{n(n+1)}}}\left(1-\frac{3(n+1)(3n+4)}{2^{\frac{3n+4}{n}}n^3(n+2)}\frac{\xi(n)}{j^{\frac{2(n+1)}{n}}}\right)~.~~~
\eeq
The field redefinition invariant expressions \eqref{a1}-\eqref{a3} are one of the main results in this note and they describe the phase structure of higher-dimensional black rings beyond the infinitely thin approximation for which the corrections proportional to $\xi(n)$ in \eqref{a1}-\eqref{a3} are not present.

\begin{figure}[H]
\centering
  \includegraphics[width=0.9\linewidth]{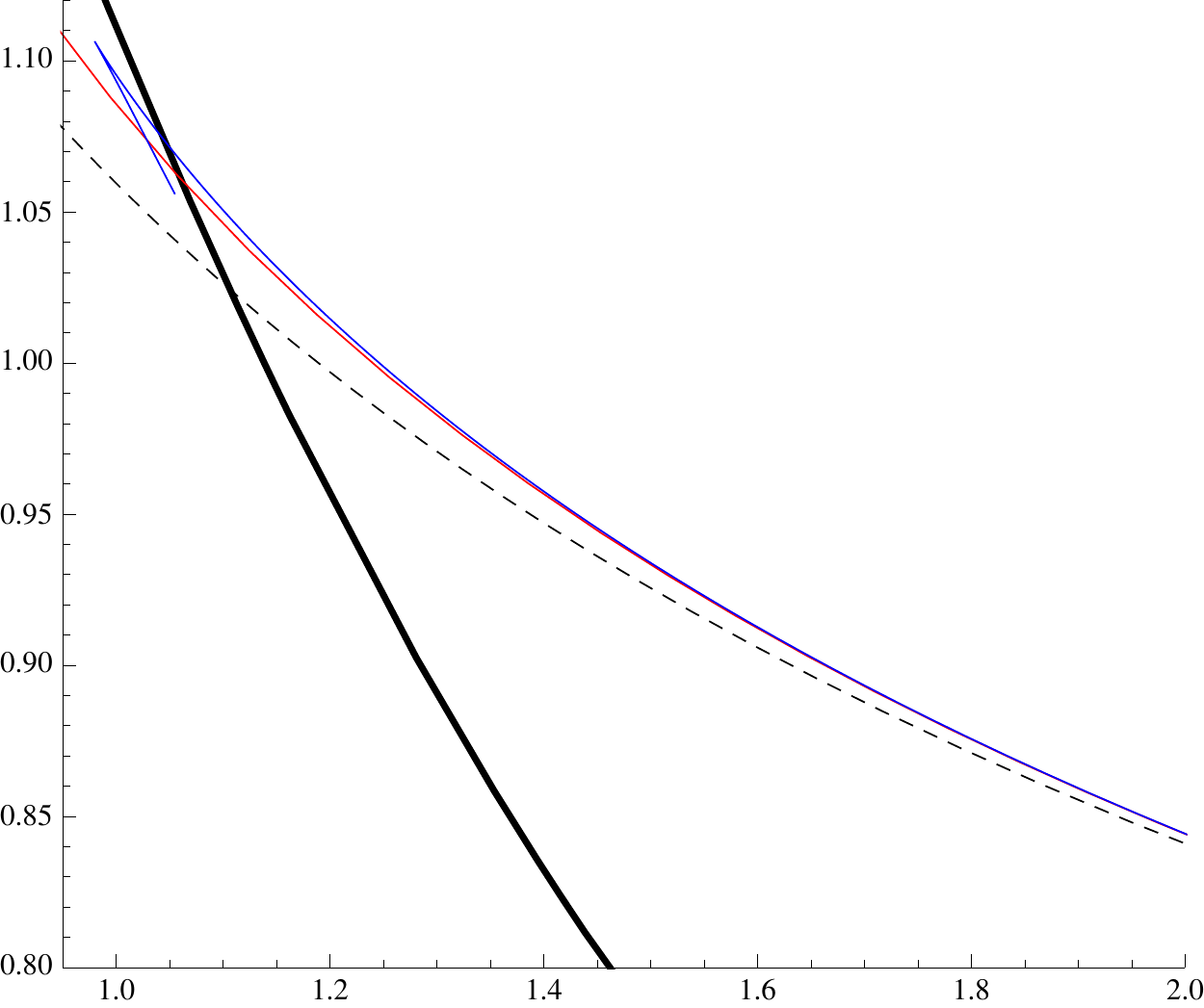} 
  \begin{picture}(0,0)(0,0)
\put(-240,160){ $ a_{\text{H}}  $}
\put(-20,-5){ $ j$}
\end{picture}	
\caption{$a_{\text{H}}$ as a function of $j$. The solid black line corresponds to the MP black hole, while the dashed black line to the black ring in the infinitely thin limit \cite{Emparan:2007wm}. The red curve is the improved phase structure given by \eqref{a1} and the blue line is the numerically obtained curve for black rings in \cite{Dias:2014cia}.}\label{avsj}
\end{figure}
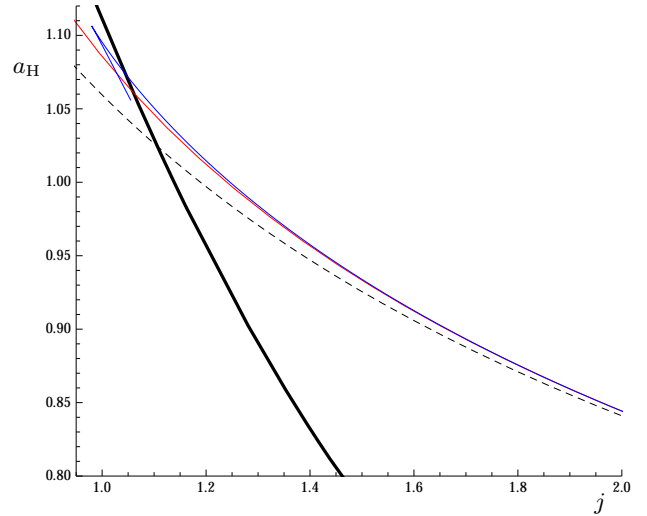

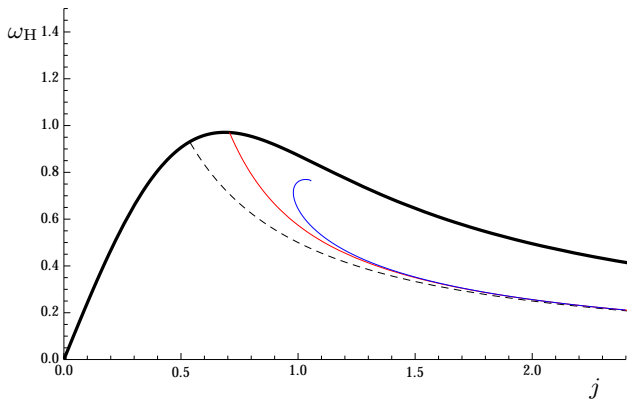
\begin{figure}[H]
\centering
  \includegraphics[width=0.9\linewidth]{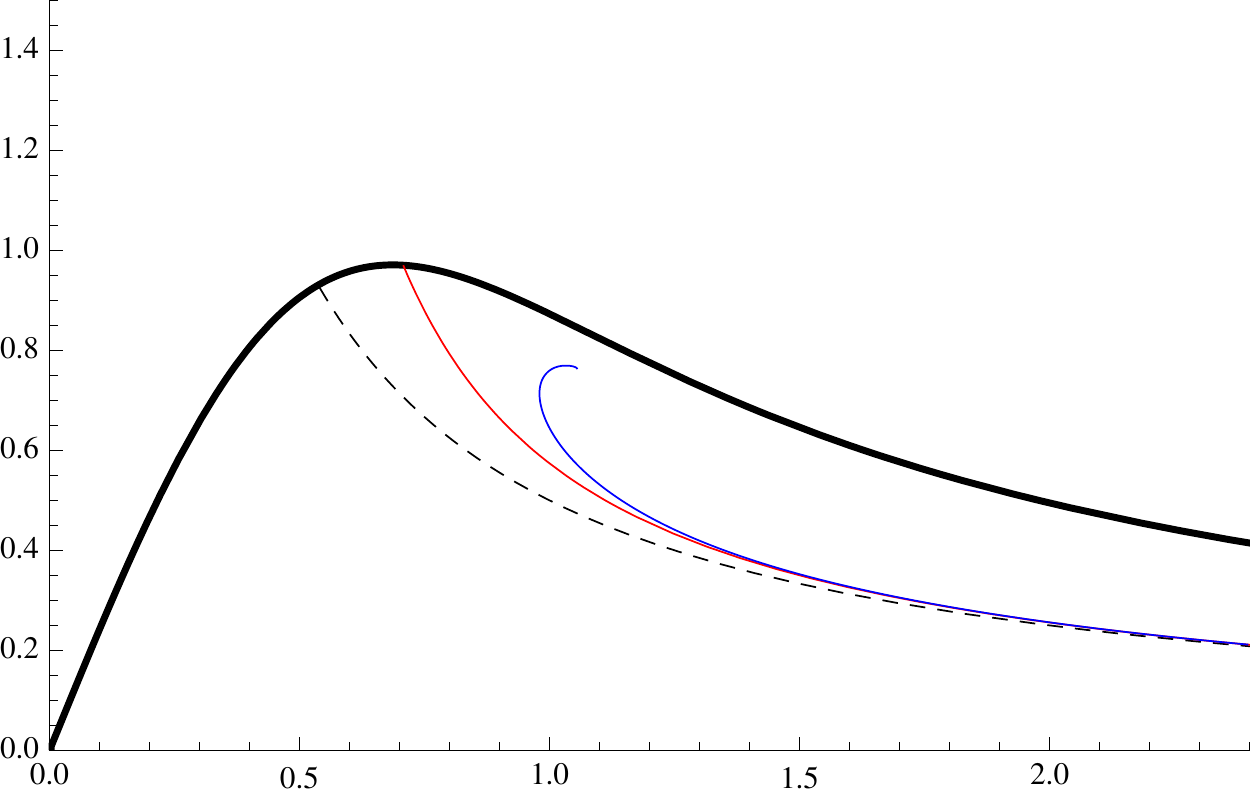} 
  \begin{picture}(0,0)(0,0)
\put(-240,130){ $ \omega_{\text{H}}  $}
\put(-20,-5){ $ j$}
\end{picture}	
\caption{$ \omega_{\text{H}}  $ as a function of $j$.}\label{wvsj}
\end{figure}

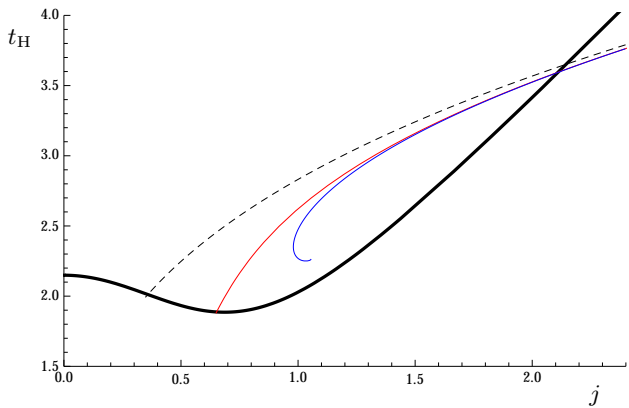
\begin{figure}
\centering
  \includegraphics[width=0.9\linewidth]{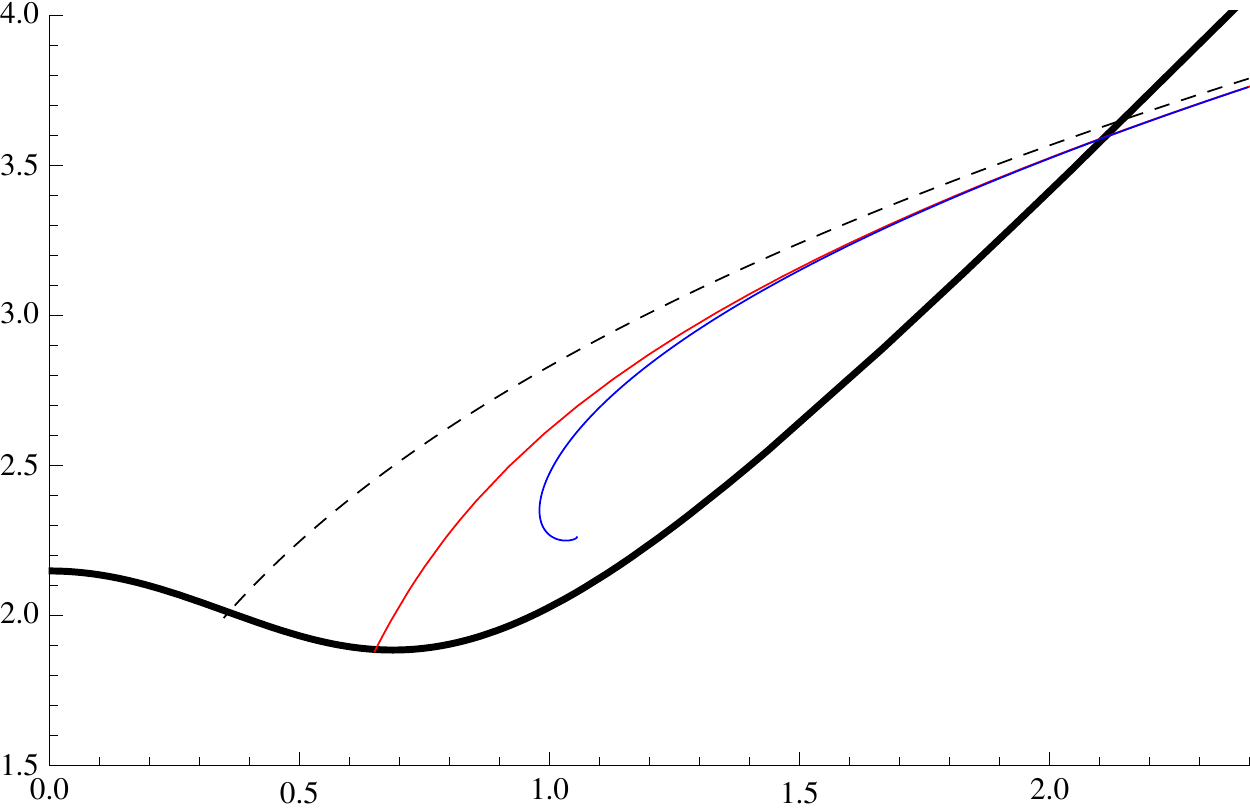} 
  \begin{picture}(0,0)(0,0)
\put(-240,130){ $ t_{\text{H}}  $}
\put(-20,-5){ $ j$}
\end{picture}	
\caption{$ t_{\text{H}}  $ as a function of $j$.}\label{tvsj}
\end{figure}

Below we compare the form of \eqref{a1}-\eqref{a3} with the infinitely thin approximation and with the numerical results recently obtained in \cite{Dias:2014cia} for $D=7$ corresponding to $n=3$. We have also plotted the corresponding quantities for singly-spinning Myers-Perry (MP) black holes in $D=7$. In Fig.~\ref{avsj} we exhibit the form of $a_{\text{H}}$ as a function of $j$, extrapolating it to values of $j\sim\mathcal{O}(1)$. We see that the red curve given by Eq.~\eqref{a1} is in striking agreement with the blue curve obtained numerically even beyond the regime of validity $\varepsilon\ll1$ of our analysis. Furthermore, we see that for a given value of the reduced angular momentum $j$ the reduced area $a_{\text{H}}$ increases compared to the infinitely thin limit. This indicates that including bending corrections in the free energy \eqref{free} leads to an increase in black hole entropy.
In Fig.~\ref{wvsj} we plot the reduced angular velocity $\omega_{\text{H}}$ as a function of $j$. We see that the effects of bending require the black hole to have a higher angular velocity for a given value of $j$ in order to compensate for the increase in the attractive force.
Finally, in Fig.~\ref{tvsj} it is shown the reduced temperature $t_{\text{H}}$ as a function of $j$. In this case the reduced temperature has decreased for a given value of $j$ compared to the infinitely thin approximation. 
The behaviour of the expressions \eqref{a1}-\eqref{a3} exhibited in Figs.~\ref{avsj}-\ref{tvsj} is generic for any $D\ge7$. It is therefore expected that the phase structure of black rings in higher dimensions is the same for any $D$ as found numerically for $D=6$ \footnote{Our analytic results cannot be compared to $D=6$ black rings because in $D=6$ back reaction corrections dominate over finite thickness corrections \cite{Armas:2011uf}.} and $D=7$ \cite{Kleihaus:2012xh, Dias:2014cia}.

\noindent\textbf{Discussion.}
In this note we have shown how the effective theory describing stationary black holes in the long-wavelength regime, also known as the blackfold approach \cite{Emparan:2009cs, Emparan:2009at}, can be used to obtain general features of higher-dimensional black rings. In such cases, the free energy of black rings \eqref{fbr} acquires an extra contribution which is the generalisation to higher codimension of the Helfrich-Canham bending energy \eqref{HC}. We see that in dimensions greater than six, the biophysical membrane effect is the leading contribution in a perturbative construction of higher-dimensional black rings, a statement which is verified for the first time by the numerical analysis of \cite{Dias:2014cia}. Furthermore, as in the case of biophysical membranes for which the effect of \eqref{HC} is to increase their surface area, we see that black rings exhibit the same phenomena, increasing their area (and hence entropy) for a given value of angular momentum.

These bending effects are generic for wide classes of black holes, in particular, for all the new black holes found in \cite{Emparan:2009vd}, the effective theory \eqref{free} governs their dynamics and it will thus provide an improved description. In this context, it would be interesting to push the perturbative constructions of \cite{Emparan:2007wm,Camps:2012hw} to next order so that the transport coefficients $\upsilon_1,\upsilon_2,\upsilon_3$ could be measured. This would also allow to derive directly Eqs.\eqref{eq1}-\eqref{eq2} from gravity as in \cite{Emparan:2007wm,Camps:2012hw}.  We note that all our analysis has resided on the measurement of the Young modulus from a first order bent metric \cite{Armas:2011uf, Camps:2012hw} and we have shown that in the case of black rings, that data is enough to predict the phase structure to second order. The same techniques can be used for charged black holes and for black holes carrying transverse spin. These cases will be addressed in a future publication \cite{almost}.

We believe that the analogy between gravity and biophysical membranes can have a fruitful development and that this opens up exciting new possibilities for applying techniques used in biophysics to black holes and vice-versa. However, we note that while the effective theory \eqref{free} is essentially the same for both black branes and biophysical membranes, their phenomenology is not since in the case of black branes the allowed topologies are further constrained by the intricate effects of stabilisation due to angular momenta. 

\noindent\textbf{Acknowledgements.}
We are grateful to \'{O}scar J. C. Dias, Jorge E. Santos and Benson Way for sharing their unpublished numerical results for 7D black rings used to plot the blue curves in Figs.\ref{avsj}-\ref{tvsj}. We also thank Joan Camps and Kentaro Tanabe for useful discussions. JA is supported by the Swiss National Science Foundation and the `Innovations- und Kooperationsprojekt C-13' of the Schweizerische Universit\"{a}tskonferenz SUK/CUS.

\end{document}